\documentstyle[12pt]{article}
\newcommand{\be}{\begin{equation}}
\newcommand{\ee}{\end{equation}}
\newcommand{\beqs}{\begin{eqnarray}}
\newcommand{\eeqs}{\end{eqnarray}}

\begin{document}
\pagestyle{plain}
\setcounter{page}{1}
\newcounter{bean}
\baselineskip16pt


\begin{titlepage}
\begin{flushright}
PUPT-1720\\
IASSNS-HEP-97/102\\
hep-th/9709108
\end{flushright}

\vspace{7 mm}

\begin{center}
{\huge Schwarzschild Black Holes in Various }

\vspace{5mm}

{\huge Dimensions from Matrix Theory }
\end{center}
\vspace{10 mm}
\begin{center}
{\large
Igor R.~Klebanov \\
}
\vspace{3mm}
Joseph Henry Laboratories\\
Princeton University\\
Princeton, New Jersey 08544\\
\vspace{3mm}
\centerline{\large and}
\vspace{3mm}
{\large Leonard Susskind\footnote
{ Permanent address:
Physics Department, Stanford University, Stanford, CA 94305.}
\\ }
\vspace{3mm}
School of Natural Sciences\\
Institute for Advanced Study\\
Princeton, New Jersey 08540
\end{center}
\vspace{7mm}
\begin{center}
{\large Abstract}
\end{center}
\noindent
In a recent paper it was shown that the properties of Schwarzschild black 
holes in $8$ dimensions are correctly described 
up to factors of order unity by
Matrix theory compactified on $T^3$.
Here we consider compactifications on tori of general dimension $d$.
Although in general little is known about the relevant $d+1$
dimensional theories on the dual tori, there are
hints from their application to near-extreme parallel Dirichlet
$d$-branes. Using these hints we get the correct mass--entropy
scaling for Schwarzschild black holes in $(11-d)$ dimensions.
\vspace{7mm}
\begin{flushleft}
September 1997

\end{flushleft}
\end{titlepage}


\newpage
\renewcommand{\baselinestretch}{1.1} 


\renewcommand{\epsilon}{\varepsilon}
\def\fixit#1{}
\def\comment#1{}
\def\equno#1{(\ref{#1})}
\def\equnos#1{(#1)}
\def\sectno#1{section~\ref{#1}}
\def\figno#1{Fig.~(\ref{#1})}
\def\D#1#2{{\partial #1 \over \partial #2}}
\def\df#1#2{{\displaystyle{#1 \over #2}}}
\def\tf#1#2{{\textstyle{#1 \over #2}}}
\def\d{{\rm d}}
\def\e{{\rm e}}
\def\i{{\rm i}}
\def\Leff{L_{\rm eff}}


\def \td {\tilde }
\def \ci {\cite}
\def \sm {$\s$-model }

\def \o {\omega}
\def \inv {^{-1}}
\def \ov {\over }
\def \four{{\textstyle{1\over 4}}}
\def \fourth{{{1\over 4}}}
\def \ha {{1\ov 2}}
\def \QQ {{\cal Q}}

\def\sqr#1#2{{\vcenter{\vbox{\hrule height.#2pt
         \hbox{\vrule width.#2pt height#1pt \kern#1pt
            \vrule width.#2pt}
         \hrule height.#2pt}}}}
\def\square{\mathop{\mathchoice\sqr34\sqr34\sqr{2.1}3\sqr{1.5}3}\nolimits}

\def\TL{\hfil$\displaystyle{##}$}
\def\TR{$\displaystyle{{}##}$\hfil}
\def\TC{\hfil$\displaystyle{##}$\hfil}
\def\TT{\hbox{##}}

\def\shortlistrefs{\footatend\bigskip\bigskip\bigskip%
Immediate\closeout\rfile\writestoppt
\baselineskip=14pt\centerline{{\bf References}}\bigskip{\frenchspacing%
\parindent=20pt\escapechar=` Input refs.tmp\vfill\eject}\nonfrenchspacing}

\def\eff{{\rm eff}}
\def\abs{{\rm abs}}
\def\hc{{\rm h.c.}}
\def\+{^\dagger}

\def\cl{{\rm cl}}

\def\M{\cal M}
\def\D#1#2{{\partial #1 \over \partial #2}}

\def\overleftrightarrow#1{\vbox{Ialign{##\crcr
     \leftrightarrow\crcr\noalign{\kern-0pt\nointerlineskip}
     $\hfil\displaystyle{#1}\hfil$\crcr}}}

\def \t {\tau}
\def \td {\tilde }
\def \ci {\cite}
\def \sm {$\s$-model }

\def \o {\omega}
\def \inv {^{-1}}
\def \ov {\over }
\def \four{{\textstyle{1\over 4}}}
\def \fourth{{{1\over 4}}}
\def \ha {{1\ov 2}}
\def \QQ {{\cal Q}}

\def \lr { \lref}
\def\np {{  Nucl. Phys. }}
\def \pl {{  Phys. Lett. }}
\def \mpl {{ Mod. Phys. Lett. }}
\def \prl {{  Phys. Rev. Lett. }}
\def \pr  {{ Phys. Rev. }}
\def \ap  {{ Ann. Phys. }}
\def \cmp {{ Commun.Math.Phys. }}
\def \ijmp {{ Int. J. Mod. Phys. }}
\def \jmp {{ J. Math. Phys.}}
\def \cqg {{ Class. Quant. Grav. }}

\section{Introduction}

In a recent paper \cite{BFKS} Schwarzschild black holes in 8 dimensions were
studied from the point of view of Matrix theory \cite{BFSS}
compactified on $T^3$.
In this particular case 
enough is known about the relevant
SYM theory on the dual torus 
to derive the properties of black holes, including the
mass--entropy relation and the physical size, up to numerical
factors of order unity. 
An important ingredient in this work was the knowledge of
the equation of state.
Certain special features of the $3+1$ dimensional SYM theory with
16 supercharges, such as its conformal invariance,
require the equation of state to be of the form,
\be \label{conf}
E\sim N^2 V_d T^4\ , \qquad S\sim N^2 V_d T^3
\ ,\ee 
where $V_d$ is the volume of the dual torus.
This equation of state is supported by the form of the near-extremal
entropy of the self-dual 3-brane found in \cite{GKP,kt}.

Here we consider generalization of
\cite{BFKS} to compactification of Matrix theory
on $T^d$. For general $d$ little is known about the 
relevant large $N$ SYM theory from
first principles. However, the same theory is supposed to describe
$N$ concident Dirichlet $d$-branes \cite{EW}. 
Thus, we will assume that the equation
of state for this theory can be read off from the 
near-extremal entropy of the RR charged classical $d$-brane
solution that was found in \cite{kt}. 
Then we will see that the same strategy that 
has worked for the $D=8$
Schwarzscild black holes \cite{BFKS} continues to work in 
other dimensions, and we find
the correct scaling of mass vs. entropy.

Let us review the strategy.
Matrix theory is best thought of as the Discretized Light-Cone
Quantization (DLCQ) of M-theory \cite{S97}, i.e. compactification
on a light-like circle of radius $R$. Accordingly, the longitudinal
momentum $P_-=P^+$ is quantized in integer multiples of
$1/R$,
\be
P_- = {N\over R}\ .
\ee
We further compactify $d$ transverse coordinates on a $d$-dimensional
torus. For simplicity, we consider this torus to be
``square'' with equal circumferences given by $L$.

The Matrix theory conjecture is that the sector of the theory with
given value of $N$ is exactly described by $U(N)$ SYM theory
in $d+1$ dimensions 
with 16 real supercharges. This theory lives on a dual torus with
circumferences \cite{LS,GRT}
\be \label{dual}
\Sigma \sim {l_{11}^3\over RL}
\ .
\ee
For physical applications $N$ has to be taken sufficiently large
to achieve desired ``resolution'' of a given system, i.e. to capture
the important degrees of freedom. In particular, a black hole of 
size $R_s$ has to be boosted to sufficient momentum that its longitudinal
dimension fits within the circle of radius $R$.
In \cite{BFKS} it was shown that this requires $N$ to be at least
of the order of the black hole entropy, $N \sim S$. This is the
regime where we will be working.

Determination of the mass--entropy relation for the
black hole goes as follows. 
We start with the relation between 
the Matrix theory energy and entropy (the equation of state)
for given $N$,
\be
E= E(N, S)\ .
\ee
Next we observe that the Matrix theory hamiltonian is identified
with the DLCQ energy according to
\be \label{energy}
E= {M^2\over P_-}= {M^2 R\over  N}
\ .\ee
Thus, we find
\be\label{massdef}
M^2 ={N\over R} E(N,S)\ .
\ee
Choosing $N\sim S$, we have
\be\label{newmassdef}
M^2 \sim {S\over R} E(S,S)\ .
\ee
Note that the Matrix hamiltonian is explicitly proportional to
$R$, so that $R$ cancels in (\ref{newmassdef}) leaving a relation between
mass and entropy.

In \cite{BFKS} it was argued that there is a sector of the 
$3+1$ dimensional SYM theory
where the equation of state (\ref{conf}) indeed holds down to entropies
of order $N$, where the temperature is very low,
\be
T\sim {1\over N^{1/3} \Sigma}
\ .
\ee
This sector may be thought of as a single 3-brane, each of whose
sides is wrapped $N^{1/3}$ times over the torus.
Using the equation of state (\ref{conf}) for $d=3$,
and applying the above strategy, it was found in \cite{BFKS} that
\be
S\sim M^{6/5} G_8^{1/5}\ ,
\ee
which is correct for $D=8$ black holes.

\section{Schwarzschild Black Holes In $D\neq 8$}

Analysis of Schwarzschild black holes in dimensions $D \neq 8$ is hampered
by the lack of understanding of the
relevant SYM theory in $d+1$ dimensions ($d= 11-D$).
Nevertheless, we can get useful information on its 
equation of state by studying near-extremal 
RR charged $d$-branes, as was done in \cite{kt}.
We will show that, if we use such an equation of
state, then
the mass--entropy relation works out correctly for 
Matrix theory Schwarzschild black holes in all dimensions.
We should emphasize that we do not have a first priciples
derivation of these equations of state and are puzzled by
some of their strange features, such as the negative specific heat
for $d>5$. Nevertheless, it is very interesting that the approach
to Schwarzschild black holes
that has worked for $D=8$ continues to work in other dimensions.

First we demonstrate a few examples, and then work out
the general formula.
For $d=1$ we conjecture the following equation of state,
\be \label{done}
S\sim N^{3/2} \Sigma T^2 g^{-1}\ ,\qquad
E\sim N^{3/2} \Sigma T^3 g^{-1}\ .
\ee
The scalings with respect to $N$, $\Sigma$ and $T$ are exactly as
for near-extremal RR charged strings \cite{kt}.
The only other quantity in the SYM theory that can be used
to make the formulae dimensionally correct is the
coupling $g$, and we have inserted it appropriately.
The power $N^{3/2}$ seems surprising, but if we follow the
usual large $N$ logic and introduce $e = g\sqrt N$ then
(\ref{done}) becomes\footnote{We are grateful to A. Polyakov
for a useful discussion on this.}
\be \label{donenew}
S\sim N^2 \Sigma T^2 e^{-1}\ ,\qquad
E\sim N^2 \Sigma T^3 e^{-1}\ ,
\ee
consistent with there being $O(N^2)$ degrees of freedom.

The equation of state (\ref{done}) is like that of
a superconformal theory in $2+1$ dimensions. Thus, we conjecture
that the $1+1$ dimensional large $N$ SYM theory with 16
supercharges develops an extra dimension of size $1/g$ (obviously,
this dimension is macroscopic only for weak coupling).
The coupling constant is given by \cite{fhrs}
\be\label{coupling}
g^2\sim {R^2\over L l_{11}^3}
\ .\ee
Using (\ref{dual})
we find that the dimensionless coupling $g \Sigma$ is independent
of $R$. Just as for $d=3$ our strategy is to work with $N\sim S$.
From (\ref{coupling}) 
and (\ref{dual}) we find
\be\label{bht}
T \sim N^{-1/4} R L^{1/4} l_{11}^{-9/4}
\ .\ee
Applying the strategy reviewed in the introduction, 
we find that the $R$ dependence cancels out, which means that
the equation of state (\ref{done}) gives results consistent with the
boost invariance.
We arrive at the following mass--entropy relation,
\be\label{meten}
S\sim M^{8/7} G_{10}^{1/7}
\ ,
\ee
which is indeed correct for $D=10$ Schwarzschild black holes.

It is important to determine the range of validity of the
equation of state (\ref{done}).\footnote{We are grateful to
E. Witten for illuminating remarks on this issue.}
We expect (\ref{done}) to hold for temperatures no greater than $T_{max}$.
For $T>T_{max}$ the theory should become approximately free, with
the equation of state,
\be \label{donefree}
S\sim N^2 \Sigma T \ ,\qquad
E\sim N^2 \Sigma T^2 \ .
\ee
This equation of state was used in \cite{HP} to
match the entropy of D1-branes sufficiently far from extremality.
Comparing with (\ref{done}), we see that the transition temperature
is $T_{max}\sim N^{1/2} g$. This is exactly where we expect it to be,
since here the large $N$ dimensionless coupling parameter,
\be 
\kappa={g^2 N\over T^2}\ ,
\ee
is of order one. For $T\gg T_{max}$, $\kappa\ll 1$ and the theory
is close to being free.
The temperature (\ref{bht}) is far
below $T_{max}$ for large $N$. Thus, we are justified in using
(\ref{done}) as the equation of state for $S\sim N$.
For $S<N$ we believe that the equation of state assumes yet another
form, which can be deduced from the expectation that $M^2$ computed from
(\ref{massdef}) is independent of $N$ and satisfies (\ref{meten}).
Thus, for $S<N$ we are led to conjecture the equation of state 
\be
E(N,S) \sim {1\over N} S^{7/4} g^{1/2} \Sigma^{-1/2}
\ .
\ee

Another instructive special case is $d=4$.
Here we conjecture the following equation of state,
\be\label{dfour}
S\sim N^3 \Sigma^4 T^5 g^2\ ,\qquad
E\sim N^3 \Sigma^4 T^6 g^2\ .
\ee
The scalings with respect to $N$, $\Sigma$ and $T$ are exactly as
for near-extremal RR charged 4-branes \cite{kt}, and we have
added a power of the SYM coupling $g$ to make the dimensions
correct. In terms of the coupling $e$ kept finite in the large $N$
limit, we have
\be\label{dfournew}
S\sim N^2 \Sigma^4 T^5 e^2\ ,\qquad
E\sim N^2 \Sigma^4 T^6 e^2\ .
\ee
The fact that in our conjectured
equations of state $S$ and $E$ scale as $N^2$
for fixed $T$, $\Sigma$ and $e$ is true for general dimension $d$.

The equation of state (\ref{dfour})
is consistent with the idea that the
theory develops an extra dimension of length $g^2$ \cite{Rozali,fhrs}.
Thus, it behaves like a superconformal theory in $5+1$ dimensions.
We believe that (\ref{dfour}) is valid down to entropies of order $N$.
Using \cite{fhrs}
\be
g^2\sim {l_{11}^6\over R L^4}
\ee
and (\ref{dual}) we find
\be\label{bhtnew}
T \sim N^{-2/5} R L^{8/5} l_{11}^{-18/5}
\ .
\ee
Proceeding in the usual way we arrive at
\be
S\sim M^{5/4} G_7^{1/4}
\ ,
\ee
which is correct for $D=7$ Schwarzschild black holes.

Now consider a general case
of Matrix theory compactified on $T^d$. 
For longitudinal
momentum $N$ we are dealing with $U(N)$ SYM theory in $d+1$
dimensions.
The same theory describes $N$ coincident Dirichlet
$d$-branes. For large $N$, we can use the near-extremal 
RR-charged $d$-brane
solution to read off
the relation between the entropy and the energy for such a theory \cite{kt},
\be \label{relation}
S \sim \sqrt N E^\lambda \Sigma^{d(1-\lambda)} g^a\ , 
\qquad 2 \lambda= {D-2\over D-4}
\ .
\ee
The power of the gauge coupling is determined by
dimensional analysis,
\be 
a={8-D\over D-4}
\ .
\ee
We also need the expression for the coupling \cite{fhrs},
\be
g^2\sim {l_{11}^{3d-6} R^{3-d}\over L^d}
\ .
\ee 
Now consider this theory for entropy of order $N$.\footnote{
For $d<3$
considerations of the range of validity, such as
those given earlier for $d=1$, show that (\ref{relation})
is valid only up to some maximum entropy, $S_{max}$.
For $S> S_{max}$ we instead find the equation of state
of a free field theory. $S_{max}$
turns out to be much greater than $N$, so that our approach is
self-consistent. }
We find from (\ref{relation}) that
\be
E\sim R S^{D-4\over D-2} \left ({L^d\over l_{11}^9 }\right )^{2\over D-2}
\ .\ee
Repeating the by now familiar steps, we find
\be
S \sim M^{D-2\over D-3} G_D^{1\over D-3}
\ ,\ee
where $G_D= l_{11}^9/L^d$.
Thus, the scaling of the entropy vs. the mass works out 
correctly for $4\leq D\leq 11$.
Note, however, that the equation of state (\ref{relation})
becomes singular for $D=4$ ($d=7$).

The cases $d=5$ and $d=6$ are also special in that they are best
described in a microcanonical ensemble. For $d=5$ the entropy
(\ref{relation}) is characteristic of dynamical strings \cite{Malda}.
Interpreting the temperature following from
(\ref{relation}) as the Hagedorn temperature of the
effective strings, we find that their tension is
\be
T_{eff}\sim {1\over N g^2}
\ .
\ee
This is consistent with the strings being instantons of
fractional instanton number $1/N$.

For $d=6$, 
(\ref{relation}) has the form characteristic of dynamical 3-branes
\cite{KT}. However, the negative specific heat encountered
here suggests that compactification of Matrix theory on $T^6$ may
involve radically new physics. Indeed, there are other indications
for this.\footnote{N. Seiberg, private communication.}

\section*{Acknowledgments}

We are grateful to C. Callan, A. Polyakov and  
E. Witten for discussions on the
equations of state and to N. Seiberg for discussions of
compactification on $T^6$. 
We also thank T. Banks and A. Rajaraman for 
discussions of compactification on $T^4$.
The work of I.R.K was supported in part by the DOE grant DE-FG02-91ER40671,
the NSF Presidential Young Investigator Award PHY-9157482, and the
James S.{} McDonnell Foundation grant No.{} 91-48.  
L.S. is grateful to IAS for hospitality.
L.S. is a Raymond and Beverly Sackler Fellow at IAS and
is supported in part by NSF grants PHY-9513835 and PHY-9219345.



\end{document}